# Defecting or not defecting: how to "read" human behavior during cooperative games by EEG measurements

**Classification: neuroscience - cognitive neuroscience**

De Vico Fallani F[*,1,2], Nicosia V[*,3], Sinatra R[*,3,4], Astolfi L[1,5], Cincotti F[1], Mattia D[1], Wilke C[6,7], Doud A[6,7], Latora V[3,4], He B[6] and Babiloni F[1,2]

[1]*NEILab, Scientific Institute for Research Hospitalization and Health Care, "Fondazione Santa Lucia", Via Ardeatina 3, 00179 Rome, Italy*

[2]*Dept. Physiology and Pharmacology, University of Rome "Sapienza", P.le Aldo Moro 5, 00185 Rome, Italy*

[3]*Complex Systems Lab - Scuola Superiore di Catania, University of Catania,Via San Nullo 5, 95123 Catania, Italy*

[4]*Department of Physics and astronomy - INFN, University of Catania, Via S. Sofia 64, 95123 Catania, Italy*

[5]*Dept. Computer Systems and Information, University of Rome "Sapienza", Via Ariosto 25,00185 Rome, Italy*

[6]*Department of Biomedical Engineering, University of Minnesota, 312 Church St., 55455 Minneapolis, USA*

[7]*Center for Neuroengineering, University of Minnesota, 312 Church St., 55455 Minneapolis, USA*

[*] These authors have equally contributed to this work.

Correspondence should be addressed to:
Prof. Fabio Babiloni
Dept. Physiology and Pharmacology, University of Rome " Sapienza",P.le Aldo Moro 5, 00185 Rome, Italy
P.le A. Moro 5, 00185, Rome, Italy
Email: fabio.babiloni@uniroma1.it
Tel: +39-3287697914
Fax: +39 0623326835

*Abstract*

Understanding the neural mechanisms responsible for human social interactions is difficult, since the brain activities of two or more individuals have to be examined simultaneously and correlated with the observed social patterns. We introduce the concept of hyper-brain network, a connectivity pattern representing at once the information flow among the cortical regions of a single brain as well as the relations among the areas of two distinct brains. Graph analysis of hyper-brain networks constructed from the EEG scanning of 26 couples of individuals playing the Iterated Prisoner's Dilemma reveals the possibility to predict non-cooperative interactions during the decision-making phase. The hyper-brain networks of two-defector couples have significantly less inter-brain links and overall higher modularity - i.e. the tendency to form two separate subgraphs - than couples playing cooperative or tit-for-tat strategies. The decision to defect can be "read" in advance by evaluating the changes of connectivity pattern in the hyper-brain network.

## Introduction

Game theory provides a mathematical framework to study decision-making processes in groups of individuals. In a game, the players adopt one among a set of possible actions (strategies), and the reward or penalty for each player crucially depends on the actions taken by all players [1]. Game theory has proven useful in the investigation of the neural basis of social interactions and social decision-making. In particular, researchers have investigated what happens in the brain of subjects involved in games where each player can choose between cooperative and non-cooperative behaviors, or between altruistic and selfish behaviors, with the aim of understanding the modification of brain activity related to the selected strategy [2].

Most of the approaches used so far to characterize brain responses during social interaction have the major limitation of measuring signals from just one player at a time. The functional connectivity between the brain activities of two interacting individuals is thus not measured directly, but inferred from independent observations subsequently aggregated by statistical models which associate observed behaviors and neural activation. In the present study, we used i) simultaneous neuroelectric recordings from two subjects, i.e. EEG hyper-scanning ii) localization of cortical activity, i.e. high-resolution EEG [22] iii) and spectral Granger causality indexes, i.e. Partial Directed Coherence (PDC) [3] to estimate, in the frequency domain, the information propagation among different cortical regions within- and between-brains. We considered one of the most common cooperation games, the Iterated Prisoner's Dilemma (IPD) [4], where each player can either defect or cooperate with the other player and might punish the opponent for previous non-cooperative behavior. A scheme of the experimental setup is provided in Figure 1. The EEG period of interest (POI) is the time interval during which both players are formulating the strategies to adopt in the next round of the game i.e. the initial decision-making phase. The resulting networks of functional connectivity estimated from the cortical activity of the two players were described by a directed weighted graph [5]. Each node corresponds to a specific cortical region - also called region of interest (ROI) - of one of the two subjects' brain. A weighted link between two ROIs indicates the degree of their interaction as estimated by the PDC. In practice, we represented the functional connectivity of the two brains altogether in the same graph: a link in the graph can be either an intra-brain or an inter-brain connection, according to the fact that it expresses the relationship between two ROIs belonging to the same brain, or between a region of one brain and a region of the other brain. We named such a graph a hyper-brain network. The obtained hyper-brain networks were analyzed using tools and measures coming from complex networks theory, such as efficiency and modularity [6].

The results obtained by analyzing 26 couples of subjects show that the structure of the hyper-brain networks corresponding to situations in which individuals play cooperatively is significantly different from cases of couples playing in a "selfish" way. Specifically, the hyper-brain network obtained from a couple of players both playing as defectors exhibits the best modular separation into two clusters corresponding to the ROIs of the two distinct brains. On the contrary, the ROIs of the two brains are more intertwined when the two players adopt cooperative or tit-for-tat strategies. We also found that the modifications of the connectivity between ROIs in the frontal and pre-frontal areas of the couple's brains are the main responsible of the structural changes discriminating collaborative from selfish behaviors. Finally, we tested the possibility to predict the outcome of a game from the structural analysis of the hyper-brain network obtained from signals recorded during the decision-making process. This suggests that EEG hyper-scanning and hyper-brain networks allow the direct observation of neural signatures of human social interactions, and might play a key role in understanding the cerebral processes generating and generated by social cooperation or competition.

## *Results*

In the Iterated Prisoner's Dilemma, a player can act as a Cooperator (C), as a Defector (D), or can adopt a Tit-for-Tat (T) strategy. Therefore, the outcome of each round, or trial, of the game can be one of the six possible combinations of the individual actions. Three of them (CC, DD, and TT) are called here "pure" strategies because the two players adopt the same action, while the others (CD, CT, and DT or equivalently DC, TC, and TD) are called "mixed" strategies. For each of the 26 pairs of players, we obtained a hyper-brain network (see details in Material and methods), for each of the six possible outcomes, and for each frequency band Theta (4-7 Hz), Alpha (8-13 Hz), Beta (14-29 Hz) and Gamma (30-40 Hz).

Figure 2 illustrates, for a representative couple of subjects, the hyper-brain networks associated to the pure strategies CC, DD, and TT, in the Alpha (8-13 Hz) frequency band. Each network consists of twelve nodes representing the six specific ROIs considered in this study for each subjects' brain. Note that the selected ROIs are the same for each player. To highlight the inter-brain connectivity, only links between the two brains are illustrated in the figure. The grand-average of the hyper-brain networks associated to all the six possible strategies is also reported in **Figure S1.**

Then, we have computed standard graph measures, such as number of edges, maximum out-degree, total network weight, maximum out-strength, efficiency and clustering coefficient [6], to characterize the networks achieved during the decision-making process. The numerical results we obtained indicate that the averages over different couples of individuals are not able to discriminate among the six possible strategies (CC, DD, TT, CD, CT, DT) due to the large standard deviations (see **Figure S2 and Table S1)**. This means that averages and standard deviations of graph metrics computed over the 26 couples do not allow for the characterization of the *typical* hyper-brain network associated with a specific strategy. In fact, the actual values of link weights can vary sensibly from couple to couple, due to differences in the time-responses of each brain and in the intensity of the EEG signals coming from different subjects. This is a well-known problem in neuroimaging. Even if the neural pattern corresponding to a given external stimulus, or to the performance of a certain task, does activate similar zones in different subjects, the amplitude of the associated EEG signals as well as the size of the activated regions and the speed of response can differ substantially from subject to subject. This evidence makes hard to find neural "fingerprints" even for simple tasks [7].

### Inter-brain connectivity discovers selfish behaviors

The novelty of this study consists in classifying different social behaviors by comparing, *for each pair of individuals*, the six hyper-brain networks relative to CC, DD, TT, CD, CT, DT strategies. For each of the 26 couples involved, we have considered the *graph efficiency E*, and computed two measures, the *divisibility D* and the *modularity Q,* which give a quantitative estimation of how well the hyper-brain network can be separated into two subsets of nodes, corresponding respectively to the network of cortical regions of the two players (see Material and methods for details). A comparison of the six values of $E$, $D$ and $Q$, obtained for each couple of players and for each frequency band, allowed successful discrimination of selfish behavior from other behaviors, as reported in the pie diagrams of Figure 3 for the Theta band. The first pie diagram shows that the 50% of cases (13 couples) presents the minimal value of efficiency in the DD hyper-brain networks, the 11.6% (3 couples) in the CC hyper-brain networks, and the 19.2% (5 couples) in the TT hyper-brain networks. The remaining 19.2% (5 couples) exhibits the lowest efficiency in mixed-strategies (CD, CT and DT) hyper-brain networks. For any frequency band, the DD connectivity pattern has the lower efficiency with respect to the other five networks in approximately the 50% of the couples. Similarly, modularity and divisibility are maximal for DD strategies in about the 75% and 62% of the couples, respectively. The statistical significance of these percentages has been tested using a $\chi^2$ test, which gives p-values ranging from 0.0001 to 0.002. These results indicate that hyper-brain networks corresponding to DD have longer paths between ROIs (lower global efficiency) and a small number of links between the two brains (high divisibility), this number being much lower than expected in a random graph with the same number of nodes and links (high modularity). Conversely, as shown in the bottom panels of Figure 3, the efficiency is maximal for TT (resp. CC) in the 30% (resp. 34%) of couples, while the modularity and the divisibility are minimal for TT and CC with similar percentages. Analogous results were observed in the all the other frequency bands (see **Figure S3**).

In other words, the relationship between the brains of two-defector couples (DD) decreases significantly (i.e. the ROIs of the two brains are better separated) with respect to two-cooperator (CC) couples or tit-for-tat couples (TT). The average Z-scores computed for the three graph measures (see Material and methods) give a clearer picture of the relations between strategies across the couples. They are reported in Figure 4, which provides a compact visualization of the results obtained for different frequency bands. As illustrated by the figure, DD hyper-brain networks are well separated from networks corresponding to other strategies. In particular, the four points relative to the DD strategy cluster together at the upper–left corner of the panel (a), indicating a relatively high divisibility and, at the same time, a relatively low efficiency with respect to the other hyper-brain networks of the same couple. In addition, the four DD points in panel (b) cluster together at the upper-right region revealing that the DD hyper-brain network modularity is usually higher than the modularity of TT or CC connectivity patterns.

**ROI relative importance**

In order to understand how different cortical regions contribute to the observed separation of the two brains in the DD networks, we performed a repeated-measure ANOVA of the node total strength $s^{in}+s^{out}$ (see **ANOVA, SI** for details).

We evaluated the dependent variable $s^{in}+s^{out}$ of each single subject within the couple (n=52) for the six specific ROIs (ANOVA's first independent variable: ROI factor) and for the three pure strategies (ANOVA's second independent variable: TASK factor). Each frequency band has been considered separately (see Supplementary Material for details). This statistical test showed that different TASKs induce statistically significant differences ($p<<0.001$) in the distribution of total strength, regardless of the ROIs, for all the frequency bands. Conversely, different ROIs exhibit statistically significant differences ($p<<0.001$) in total strength, for all the frequency bands, except Alpha. The interaction of the two factors, instead, produces statistically significant differences ($p<<0.001$) only in the Beta and in the Gamma band (see **Table S2**).

Consequently, a post-hoc analysis has been performed only in the Beta and Gamma bands, in order to highlight the different role of each ROIs with respect to the different strategies. In the Beta band (see Figure 5), the Brodmann area 10_L showed a significant lower value of total strength for the DD strategy with respect to both the CC (post-hoc test $p=0.000153$) and the TT (post-hoc test $p=0.000002$) strategies. The Brodmann area 10_R showed a significant lower value of total strength for the DD strategy with respect to both the CC (post-hoc test $p=0.000004$) and the TT (post-hoc test $p=0.000002$) strategies. Moreover, the value of strength of the Brodmann area 10_R in the CC strategy is also a significant lower (post-hoc test $p=0.000011$) than in the TT strategy. The Anterior cingulate cortex ACC exhibited a significant lower value of strength for the DD strategy with respect to both the CC (post-hoc test $p=0.000002$) and the TT (post-hoc test $p=0.000002$) strategy. Differences between DD strategy and either CC or TT strategies for the other three ROIs (i.e. the Cingulate Motor Area CMA, the Brodmann areas 7_L and 7_R) are not statistically significant (post-hoc test, $p>0.001$).

Similar results were observed also for the Gamma band in **Figure S4**. Summing up, we found that the observed differences in the total strength for the DD strategy with respect to either CC or TT strategies are correlated to changes in the activity of frontal and pre-frontal areas. This result is in agreement with previous works showing that frontal and pre-frontal areas play a role in decision-making processes [10-12].

**On–line classification**

Hyper–brain networks corresponding to a given couple's DD strategy have peculiar topological features, such as lower efficiency, higher divisibility and higher modularity with respect to hyper–brains corresponding to the other strategies of the same couple.

Such differences can be exploited in order to make predictions on the strategy that a player is going to adopt, based on the on-line analysis of hyper–brain networks constructed from data recorded in the decision-making process. For each frequency band, we have implemented a non–linear classifier, more specifically a Multi-Layer Perceptron (see **Multi–layer perceptron, SI** and **Figure S5** for details), using 21 couples for training (6 networks per couple, each graph corresponding to one of the 6 different strategies, for a total number of 126 networks), and the remaining 5 couples (30 networks in total) for validation. The classification is based on the values of the Z-scores of efficiency, divisibility and modularity, and not on the actual values of the measures themselves. In fact, for each couple, the Z-score of a graph measure provides its deviation from the average value computed over all hyper-brains of the same couple. The accuracies obtained by the classifiers during validation process, i.e. the number of hyper-brain networks classified correctly as DD or non–DD out of the 30 validation patterns, are respectively: 27, 22, 26, 24 for the Theta, Alpha, Beta and Gamma frequency band.

## *Discussion*

Neuroimaging techniques have recently provided strong evidence of a close link between mind and brain. It is well known that the action of concentrating on a specific object or performing a given sensory, cognitive or motor task is reflected in different patterns of brain activity.

However, it is not clear whether the decoding of mental states, or *brain reading* [8,9], i.e. inferring what an individual is thinking from his brain activity, can be practically achieved with current neuroimaging methods. The task becomes even harder if one wants to identify neural patterns corresponding to social interactions, such as the choice to cooperate or to defect in the Iterated Prisoner's Dilemma. Results obtained in this work show quantitatively that the non-cooperative behavior of a couple of players is usually associated with peculiar brain connectivity patterns, and in general with a much lower interaction between the activities of the cortical areas of the two players. The DD hyper-brain network is radically different from the other pure strategies (CC and TT), in which the selected cortical regions of the two players are highly interconnected. In fact, there are only a few inter-brain links in the DD case, this giving simultaneously a "picture" and a physical interpretation of the selfish behavior of the subjects. Each player in the couple tends to maximize his own outcome and to minimize at the same time the opponent's outcome. This evidence is coded in the hyper-brain network: cooperation requires areas corresponding to the two brains to be intermingled, while cortical areas of selfish players are almost uncoupled.

This outcome indicates the possibility of "reading" mental states, and inferring social behavior from the brain activity of couples of individuals. In particular, these results suggest that:

   *i)* with current neuroimaging techniques, it is possible to estimate in healthy subjects patterns of functional connectivity between cortical areas, which are active in decision-making processes. In the specific case of cooperation or defection strategies in social games, such patterns appear to be linked to the decisions that were made successively by the subjects, and cannot be confused with normal cerebral activity. That is because the operative conditions for the subjects are unchanged during all the experiment.

   *ii)* the patterns of functional connectivity among cortical areas sub-serving the decision of cooperating or defecting, estimated from data recorded in the decision-making process, produce different hyper-brain networks for different observed outcomes of the game. In particular, for all the frequency bands analyzed, the level of connectivity between the ROIs of the two brains significantly decreases in the case of DD strategies, while hyper-brain networks of TT and CC trials are more tightly connected and intermingled.

*iii)* the cortical areas that are more responsible for the observed inter-connectivity decrease are located principally in the pre-frontal cortex (Brodmann areas 10 and ACC). This is particularly evident in the Beta and Gamma frequency bands (13–40 Hz), where the regions of the pre-frontal cortex of the investigated subjects contained local task-specific representation of the intended decisions before such decisions were shown on the screen. These latter results are in agreement with previous literature showing that the brain activity increases in the front-polar, lateral, medial, and pre–frontal cortex during the performance of cognitive activities such as free task selection [10], formation of intentions [11] and multitasking [12]. In fact, it has been demonstrated that in humans, a network of brain regions including not only lateral but also medial pre–frontal cortex, contains such task-specific representations [8,9]. Our results are also consistent with previous studies on isolated brains that indicated the ACC as the cortical site in which humans represent the other's intentions in the brain (theory of mind [13]). In particular, several lines of evidence have suggested the role of the ACC in effort-related decision-making, including ERP investigations [14], or lesion studies in animals [15]. Moreover, it has been proposed that ACC activity might reflect the amount of effort associated with cognitive processing in conflict monitoring [16].

In the present study, a simultaneous recording of the brain activity of two individuals (hyper-scanning) involved in the iterated prisoner's dilemma game has been obtained in the neuroelectrical domain. Before this experiment, researchers have attempted to perform hemodynamic hyper-scanning during simple games, movie observation or economic transactions [17,18]. Since EEG recordings provide high temporal resolution as compared to hemodynamic measurements, they can be used in real-time for the construction of the hyper-brain networks, the relative computation of graph measures, and the on-line prediction of the outcome for each trial of the game. In particular, all the parameters needed for source reconstruction, signal ROI estimation and PDC computing can be obtained before the actual EEG session. For instance, they could be obtained in a training session during which the players learn how to play the game, or during a rest condition where the two players are exposed to the same environment that they will experience later. In such a way, all the computations can be reduced basically to a sequence of matrices multiplication. We have verified that this processing chain requires less than 3 second on a standard single-processor computer, which makes the whole process sufficiently fast to be performed on-line according to the timing proposed in the study. In addition, the multi-layer perceptron classification can be performed in real time, since it only requires few milliseconds (see **Multi–layer perceptron, SI** for details).

The results presented here indicate that a non-linear classifier is able to discriminate the DD strategy with up to 90% of accuracy. Therefore, the proposed classification process is able to predict the defection strategy of the two players *before* they press the keyboard buttons to communicate their choices. In principle, a similar approach can be used to train non-linear classifiers to predict CC and TT strategies as well. Such an extension would probably require only a larger dataset, i.e. more than 26 couples.

In conclusion, we have presented an application of complex network theory to the analysis of functional brain connectivity and to the study of its correlation with observed social behaviors. Indeed, many of the theoretical results obtained in the last few years in the field of complex networks are still waiting to be exploited in the field of neuroscience, and could potentially give us a better insight into the structure and meaning of complex biological systems, as they have already done with social and technological networks. The fact that graph theoretical indexes can also be used to better understand how the human brain works [19,20,21,29], suggests that hyper-brain networks can be adopted in the near future as a valuable reference model for further investigations of the mechanisms that are the bases of social empathy [28].

## Methodological limitations

**Hyper-brain network size.** The most relevant limitation of the present study is the selection of a relatively small number of ROIs (i.e. 12 ROIs in total, six for each subject) to obtain hyper-brain networks. This aspect limits the power of the graph theoretical approach and restricts the cortical networks to a subset of predefined ROIs. This limitation is mainly due to the Partial Directed Coherence index (PDC), which is based on the generation of a valid multivariate autoregressive model (MVAR) from the estimated cortical time series. The PDC is one of the most powerful methods to reveal directed information flows between time series. Other simple methods, like spectral coherence [30], are currently available in the literature. Those bivariate methods allow the estimation of the functional connectivity among a larger number of cortical signals. Nevertheless, the advantage of MVAR models with respect to other standard bivariate methods have been already demonstrated [31]. In fact, they can efficiently detect and remove the statistical spurious links from the functional connectivity estimation, even if in most cases, like PDC, MVAR methods need a large amount of data to obtain a reliable connectivity pattern. The precision in the estimation of parameters using MVAR models requires an appropriate length of EEG recordings, as the number of such parameters substantially grows when the number of time series to be modelled (i.e. the number of nodes in the final network) increases. Other studies have already shown that the precision of the connectivity estimation by MVAR models is highly sensitive to the length of the gathered EEG data [23, 32]. In practical cases, in which a limited amount of EEG data are available, the size of the MVAR model does not allow to take into account more than 12-15 times series at a time. In the present study, given the amount of collected EEG data, a maximum of twelve cortical areas could be accurately modeled by the PDC.

**ROIs selection.** The selection of the six cortical areas considered in the present study was based on a list of ROIs included in the so called Parieto-Frontal Integration Theory (P-FIT [33]), which describes the regions of the human brain involved intelligence and reasoning tasks. This theoretical model (P-FIT) includes the dorsolateral prefrontal cortex (i.e. Brodmann areas 10), the superior (Brodmann area 7) parietal lobule, and the anterior cingulate (Brodmann area 32, i.e. ACC). In addition, there is evidence that the most caudal region of the medial frontal cortex containing cingulate motor areas (i.e. CMA) is involved in movements of the hands and other body parts ([34]). In particular, the activity in the regions including CMA has been also related directly to behavioural response rate ([35]). Since at the end of the experimental task subjects had to press a button to make their decision, the inclusion of one of the cortical stations related to movement could be of interest to understand its possible role within the functional connectivity pattern.

## *Material and methods*

### Iterated Prisoner's Dilemma

Each couple of subjects plays repeatedly a Prisoner's Dilemma game. In the Prisoner's Dilemma game, a player can choose one of the two possible strategies: either to cooperate (C) with the other player or to defect (D). Consequently, the possible outcomes of the games are four: both players cooperate (CC), first player cooperates while the other defects (CD), first player defects while the second cooperates (DC), both players defect (DD). We adopted the following payoff matrix:

$$P = \begin{pmatrix} P^{CC} & P^{CD} \\ P^{DC} & P^{DD} \end{pmatrix} = \begin{pmatrix} 2 & 0 \\ 3 & 1 \end{pmatrix} \quad (1)$$

for the payoff Π received by the first player in the four outcomes. For the second player the transpose matrix is used. When the Prisoner's Dilemma is played iteratively the situation is more complicated, since a player remembers previous actions of the opponent and can change the strategy accordingly. Here, we classify three possible strategies for a player in each trial (as in **Figure S6**): i) cooperative strategy (Cop), when a player who is playing defection, starts to cooperate as soon as the other player defects, or when a player who is playing cooperation, continues to do so for all the possible actions of the opponent; ii) defector strategy (Dft), when a player who is playing cooperation, starts to defect as soon as the other player cooperates, or when a player who is playing defection, continues to do so for all the possible actions of the opponent; iii) tit-for-tat strategy (Tft), when a player who is cooperating switches to defection if the opponent defects, or when a player who is defecting switches to cooperation if the opponent cooperates.

**Experimental design**

The experiments were conducted by the Neuroelectrical Imaging and Brain Computer Interface laboratory (NEILab) at the Scientific Institute for Research, Hospitalization and Health Care, "Fondazione Santa Lucia" in Rome (Italy) and by the Department of Biomedical Engineering in Minneapolis (USA). All the subjects involved in the experiment were recruited by advertisement. Written informed consent was obtained from each subject after the explanation of the study, which was approved by the local institutional ethics committee of the Scientific Institute for Research, Hospitalization and Health Care, "Fondazione Santa Lucia" in Rome and by the Institutional Review Board of the University of Minnesota. Fifty-two voluntary and healthy subjects (age ranging from 23 to 33 years) participated in our experiment. They had no history of neurological or psychiatric disorders and they were free from medications, alcohol, or drug abuse. In the experimental setup, each of the 26 couples of subjects played the iterated Prisoner's Dilemma game of at least 200 trials. Every round, or trial, players were asked to choose either to cooperate (C) or defect (D) through a special keyboard. A trial (k) consists of two distinct time intervals. During the first interval, players have to communicate their strategies on the base of the outcome at the previous trial (k-1). Typically, this interval ranged from 0.5 seconds to 2 seconds. After communicating their choice, a report summarizing the strategy and the score at the trial (k) is displayed for 4 seconds. At the beginning of this second interval, the two subjects make the new decision to be communicated in the next trial (k+1). For the subsequent off-line analysis, we considered the first second (i.e., 1 s of EEG recordings) as period of interest (POI) for the initial decision-making processes. All the choices of the iterated game were stored and subsequently used to classify trials, as described previously. All the subjects involved in this experiment were asked to play several games before the EEG recording session. In this way, we tried to reduce the confounding effects related to the novelty of the task, and make the subjects confident about the choices, the type of visual outputs and the possible emotions they could have experienced during a real experimental session. To obtain a good alternation of choices during the game, and to avoid any maintenance effects that could also affect the decision processes, all the subjects were strongly encouraged to avoid systematic behaviors throughout the whole game.

**EEG recordings and cortical activity**

For the EEG data acquisitions, the participants were comfortably seated on a reclining chair in an electrically shielded and dimly lit room. Two separate 64-channel systems (BrainAmp, Brainproducts GmbH, Germany) were used to record EEG signals at a sampling frequency of 200Hz. In accordance to an extension of the 10-20 international system, each electrode cap was composed of 64 sensors. The three dimensional electrode positions were obtained by using a photogrammetric localization (Photomodeler, Eos Systems Inc., Canada) with respect to the anatomic landmarks: nasion and the two pre-auricular marks. The first step of the off-line analysis consisted in band-pass filtering (0.1-45 Hz) the recorded EEG signals to allow the electrophysiology technicians to better recognize and remove the trials affected by the noise in the frequency band of interest.

Cortical activity from scalp EEG recordings was estimated by using an average realistic head model (MNI template, http://www.loni.ucla.edu/ICBM/) consisting of four concentric surfaces: scalp, inner skull, outer skull and cortex. Each surface is composed of approximately 3000 uniformly disposed vertices, each corresponding to one current dipole. The estimation of the current density strength in six regions of interest (ROIs) was obtained by solving the electromagnetic linear inverse problem according to [22] (see also **High-resolution EEG, SI).** Namely, they are the Brodmann area 10_L for the left hemisphere and 10_R for the right hemisphere, the Anterior Cingulate Cortex (ACC), the Cingulate Motor Area (CMA), the Brodmann area 7_L for the left hemisphere and 7_R for the right one.

Finally, the cortical signals of the single players in each trial were classified as Cooperation (C), Defection (D), or as Tit-for-Tat (T) according to the rules specified above. Thus, three different subsets of trials C, D and T were collected for each subject.

## Hyper-brain network

A merged dataset was constructed by considering data from the six cortical regions of the two subjects, thus yielding a set of 12 cortical signals. In order to clean out existing differences in the average activity and variance of the signals of two distinct subjects' brains, a z-transformation of the original cortical waveforms has been taken into account (see **Hyper-brain, SI** for details). The transformed cortical waveforms, one for each ROI of the merged data set, were then processed to estimate functional connectivity by means of the Partial Directed Coherence (PDC) method [23]. The PDC is a multivariate spectral measure used to determine the directed influences between any given pair of signals in a multivariate data set (see also **MVAR models, SI**). In order to consider only the functional links that are not due to chance, we used a statistical significance threshold referred to as the spectral causality criterion (SCC) [3]. According to this approach, only connections whose intensities are higher than the predetermined threshold are taken into account, while the remaining connections are excluded. Concerning the spectral properties of the EEG signals, we selected four frequency bands of interest (Theta 4-7 Hz, Alpha 8-12 Hz, Beta 13-29 Hz and Gamma 30-40 Hz) and we gathered the corresponding cortical networks by averaging the values within the respective range. Finally, for each band, we produce six different graphs, respectively corresponding to CC, DD, TT, CD, CT, DT cases. Each graph has N = 12 nodes. The first six nodes of the graph correspond to ROIs of the first player, while the remaining 6 nodes correspond to the ROIs of the second player. We call these graphs *hyper-brain networks*, since they represent at the same time the correlations between ROIs in the same brain and correlations across the two brains.

## Graph indexes

A directed weighted graph of *N* nodes can be represented by a *NxN* weighted adjacency matrix $W=\{w_{ij}\}$, where $w_{ij}>0$ is the weight associated to the directed arc from node *i* to node *j*, and in general $w_{ij} \neq w_{ji}$. The most intuitive index of a graph is its total number of links. This value measures the overall level of connectivity within the system. The respective weighted version is the total network weight that is the sum of all arc weights in the graph. The out- and in-degree of node *i* are defined respectively as the number of out- and in-going arcs. The sum of weights of the out- and in-coming arcs of a node *i* are called out- and in-strength.

**Efficiency.** The performance of a network can be measured by assuming that information flows along shortest paths and that the efficiency in the communication between two nodes *i* and *j* is inversely proportional to their shortest distance $d_{ij}$, i.e. the smallest sum of arc weights of all possible paths from *i* to *j*. Namely, the *efficiency* index *E* of a graph, is defined as [24]:

$$E = \frac{1}{N(N-1)} \sum_{i \neq j=1}^{N} \frac{1}{d_{ij}} \qquad (2)$$

If there is no path from *i* to *j*, $d_{ij}=\infty$ and the couple *(i,j)* does not contribute to the graph efficiency. Large distances imply small efficiency, while short distances imply high efficiency, with the efficiency being maximal in a fully connected graph.

**Divisibility and Modularity.** We have also implemented two measures to quantify how well the graph *G* can be divided into two sets of nodes $B_1$ and $B_2$, corresponding to the brains of the two players.

*Divisibility D* is defined as:

$$D = \frac{W}{\sum w_{ij}[1 - \delta(C_i, C_j)] + k} \qquad (3)$$

where $C_i$ indicates the community to which the node *I* belongs (here we can have either $C_i=B_1$ or $C_i=B_2$); the $\delta$ function yields *1* if vertices *i* and *j* are in the same community (i.e. in the same brain), and 0 otherwise; *k* is a positive constant (here set equal to *W*) to avoid possible divergence of *D*. The divisibility *D* is actually the inverse of the cut size [25] extended to weighted graphs.

M*odularity Q*, originally defined for unweighted graphs [26], measures the difference between the fraction of arcs connecting nodes belonging to the same community in the actual graph and its expected value in a random graph. Modularity *Q* in the case of directed weighted graphs reads [27]:

$$Q = \frac{1}{W} \sum_{ij} \left( w_{ij} - \frac{s_{i,out} s_{j,in}}{W} \right) \delta(C_i, C_j) \qquad (4)$$

where the $\delta$ function yields *1* if vertices *i* and *j* are in the same community *C* (here in the same brain, $B_1$ or $B_2$), and *0* otherwise, as in the case of divisibility *D*. As a result, in the expression of *Q*, the only contributions come from couples of nodes belonging to the same brain. Hence, the higher is the value of modularity, the better is the partition of the networks into the two communities $B_1$ and $B_2$.

In order to compare network measures for different strategies $\tau$ ( $\tau = CC, DD, TT, CD, CT, DT$ ) of the same couple *k* (*k*= 1, ..., 26), we introduce the Z-score, $Z_\tau^k(x)$, of a generic network measure *x* (*x* being the efficiency *E*, the divisibility *D*, or the modularity *Q*) as:

$$Z_\tau^k(x) = \frac{x_\tau^k - \underline{x}^k}{\sigma^k} \qquad (5)$$

The averages $\underline{x}^k$ and the standard deviations $\sigma^k$, are evaluated, for each value of *k*, over all strategies $\tau$. Finally, the average Z-score, $\langle Z_\tau(x) \rangle$, is evaluated, for each strategy $\tau$, by averaging the Z-scores, $Z_\tau^k(x)$, over all couples *k*:

$$\langle Z_t(x) \rangle = \left\langle \frac{x_t^k - \langle x^k \rangle}{s^k} \right\rangle_k \qquad (6)$$

## *Figure legends*

**Figure 1. Timeline of the experiment.**
At each round, or trial, players are asked to choose either to cooperate (C) or defect (D) through a special keyboard. A trial (k) consists of two distinct time intervals. During the first interval, players have to communicate their strategies on the base of the outcome at the previous trial (k-1). Typically, this interval ranged from 0.5 seconds to 2 seconds. After communicating their choice, a report summarizing the strategy and the score at the trial (k) is displayed for 4 seconds. At the beginning of this second interval, the two subjects make the new decision to be communicated in the next trial (k+1). In particular, we considered the first second (i.e., 1 s of EEG recordings) as period of interest (POI) for the initial decision-making processes.

**Figure 2. Inter-brain connectivity for pure strategies in the Alpha band.**
Two generic players are represented by the realistic head models used to estimate the cortical activity in the same six regions of interest (ROIs). Different colored points indicate the barycenters of these ROIs on the semi-transparent cortex. For the sake of simplicity, we didn't label the ROIs of each subplot, but just two for the CC (7_L, 10_L), TT (7_R, 10_R) and DD (CMA, ACC) subplot. Only links between the two brains are illustrated in each hyper-brain network, i.e. the inter-brain connections. The size and the color of each directed connection represent the PDC values of a representative couples of subjects in the Alpha (8-13 Hz) frequency band.

**Figure 3: Pie diagrams of efficiency *E*, divisibility *D* and modularity *Q* in the Theta band.**
Top panels: from left to right the diagrams represent the percentage of cases - over the 26 couples - in which graph efficiency *E* is minimal, whilst the divisibility *D* and modularity *Q* are maximal. Bottom panels: percentage of cases - over the 26 couples - in which *E* is maximal and *D* and *Q* are minimal. Blue areas represent pure cooperation CC, red areas represent pure defection DD, green areas represent pure tit-for-tat TT. Mixed situations CD, CT, and DT are represented by white areas. The results are reported for the Theta band (4-7 Hz). Similar pie diagrams for the other frequency bands are in **Figure S3**.

**Figure 4. Scatter plot of efficiency E, divisibility D and modularity Q during cooperation (CC), defection (DD) and tit-for-tat (TT).**
For each couple *x*, and each strategy $\tau$, the Z-scores are computed as in formula 5 (Methods Section). Then $\langle Z_\tau(x) \rangle$ is evaluated as an average of $Z_\tau^k(x)$ over all the 26 couples. For each strategy, and each frequency band, we report in panel (a), the average Z–score for the measure of divisibility, $\langle Z_\tau(D) \rangle$, vs. the average Z–score of the efficiency, $\langle Z_\tau(E) \rangle$, and in panel (b), the average Z–score of divisibility, $\langle Z_\tau(D) \rangle$, vs. the average Z–score of the modularity, $\langle Z_\tau(Q) \rangle$. Red squares represent DD values; blue circles represent CC values and green diamonds TT values. The Greek letter beside each symbol indicates the considered frequency band.

**Figure 5. Average values of total strength $s^{in}+s^{out}$ of the ROIs during cooperation (CC), defection (DD) and tit-for-tat (TT) in the Beta band (14-29 Hz).**
Values of total strength (y-axis) are obtained by considering the ROIs (x-axis) of each single subject within the couple. Thus, they represent the average of 52 subjects, i.e. 26 couples. Each line corresponds to a different task: CC (blue circles), DD (red squares) and TT (green diamonds). Vertical bars denote 0.95 confidence intervals. Single stars indicate the ROI where the DD strategy is significantly different (p<0.001) from the CC and from the TT strategy. A double star marks the ROI where all the three strategies are significantly different (p<0.001).

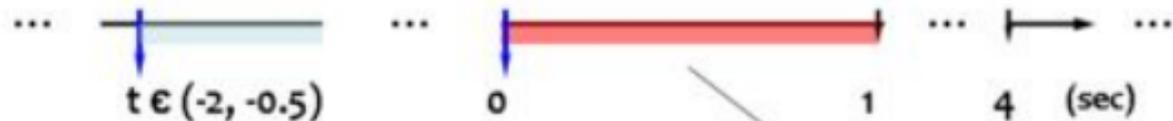

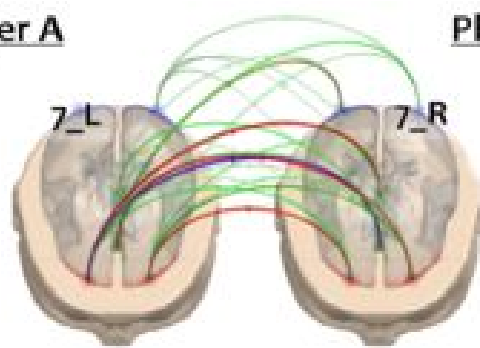
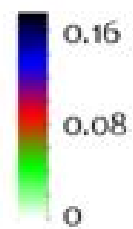
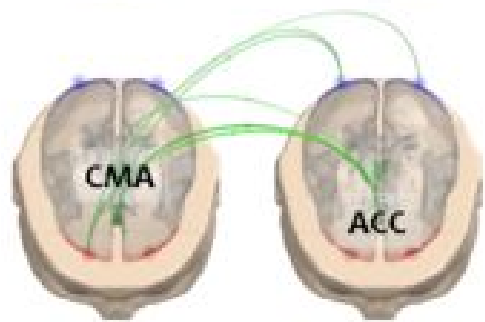
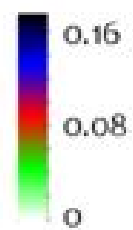
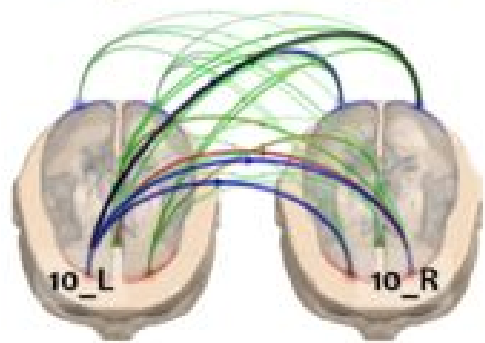
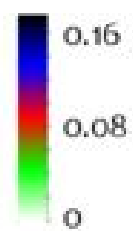

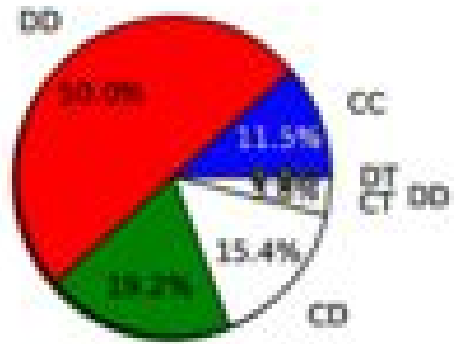 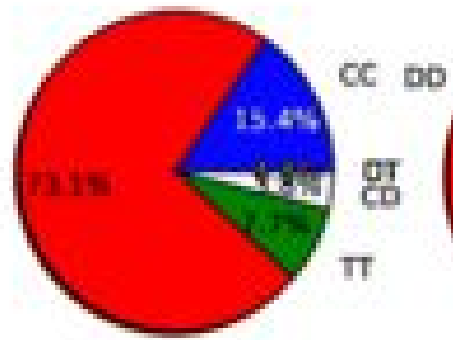 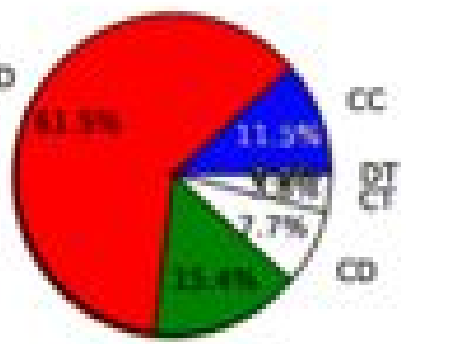
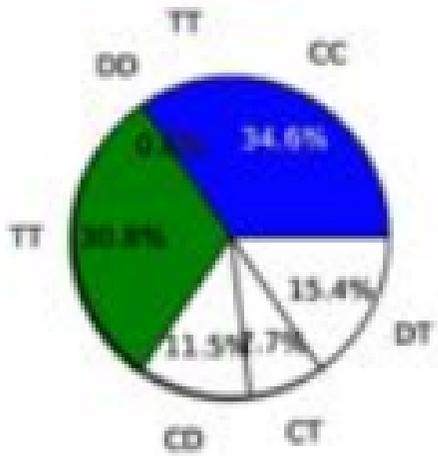 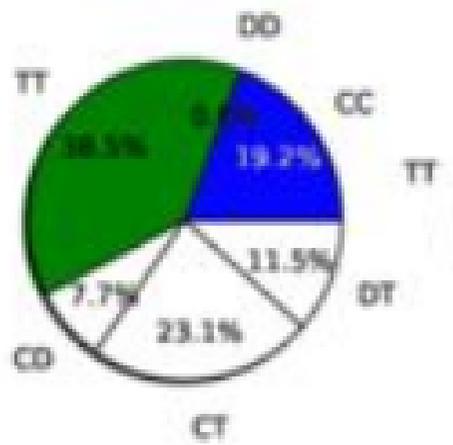 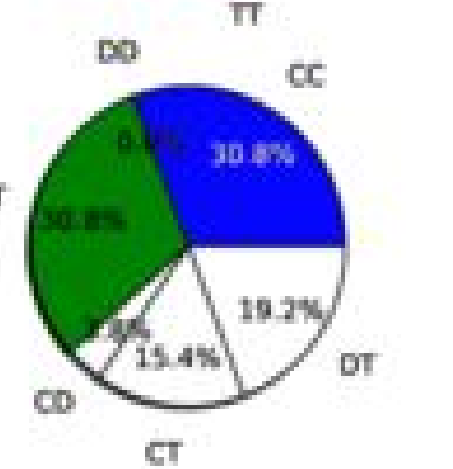

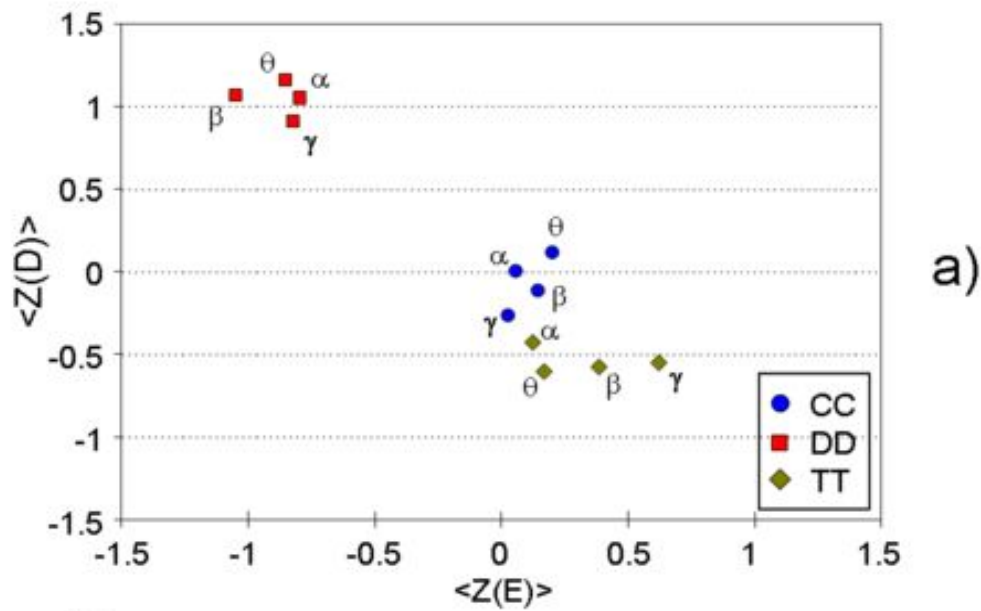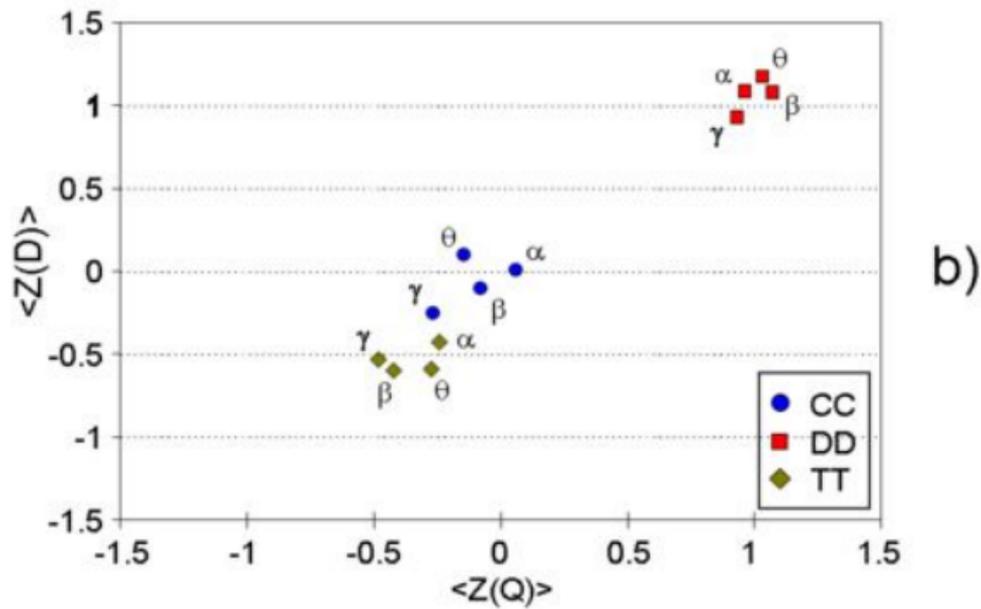

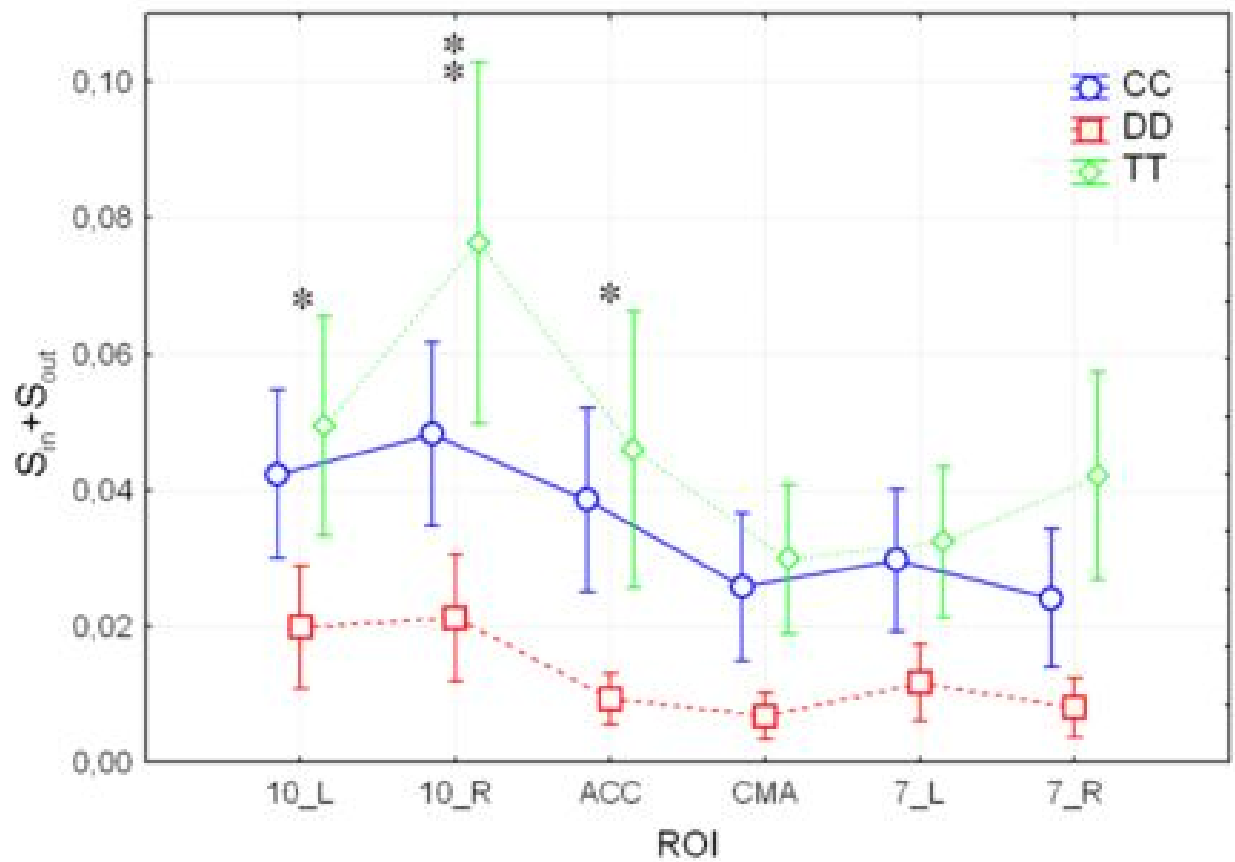